\documentclass[pre,showpacs,floatfix,twocolumn]{revtex4}
\usepackage{graphicx}
\usepackage{amssymb}
\usepackage{dcolumn}
\usepackage{bm}

\usepackage{color}

\begin{document}

\title{Realistic spatial and temporal earthquake distributions in a modified
Olami-Feder-Christensen model}

\author{E. A. Jagla}

\affiliation{Centro At\'omico Bariloche, Comisi\'on Nacional de Energ\'{\i}a At\'omica, 
(8400) Bariloche, Argentina}

\begin{abstract}

The Olami-Feder-Christensen model describes a limiting case of an elastic surface that slides on top of 
a substrate, and is one of the simplest models that display
some features observed in actual seismicity patterns. However, temporal and 
spatial correlation of real earthquakes are not correctly described by this model in its original form.
I propose and study a modified version of the model, that includes a mechanism of structural relaxation.
With this modification, realistic features of seismicity emerge, that are not obtained with the original version, mainly: aftershocks that cluster spatially around the slip surface of the main shock and follow the Omori law, and averaged frictional properties similar to those observed in rock friction, in particular the velocity weakening effect.
In addition, a Gutenberg-Richter (GR) law for the decaying of number of earthquakes with magnitude is obtained,
with a decaying exponent within the range of experimentally observed values. Contrary to the original version of the model, a realistic value of the exponent appears without the necessity to fine tune any parameter.

\end{abstract}
\maketitle

\section{Introduction}
The interest to describe the seismic phenomenon as originated in instabilities of dynamical systems 
has steadily increased in the last years. Although simplistic models cannot expect to reproduce the
full phenomenology observed in real seismicity, it has become clear that a relatively fair understanding
of some prominent features can be obtained using rather simple models. Historically, the first model of this type is the one proposed by Burridge and Knopoff (BK)\cite{bk}. 
They considered a chain of elastically interacting rigid blocks (assumed to model portions of a tectonic plate) that are forced to slide onto an underlying surface. 
In order to obtain instabilities during motion that can be associated to earthquakes, a crucial ingredient in this model is the use of a `velocity weakening' friction force between blocks and substrate, i.e., a friction force that decreases as the relative velocity increases. This was shown to generate instabilities that
produce abrupt and potentially large rearrangements of the blocks (the `earthquakes')\cite{langer}. The model is considered to be a paradigmatic case of self-organized criticality\cite{bak,bak2}, since the number of earthquakes $N$ within a fixed magnitude interval decays (albeit in a limited magnitude range) exponentially 
with the magnitude $M$ of the events, reproducing the empirical Gutenberg-Richter law\cite{gr}, namely $N(M)\sim 10^{-bM}$. For actual earthquakes the exponent $b$ is usually found to be close to 1. 

A number of modifications and generalizations have been proposed to this model over the years.
I will concentrate in the work of 
Olami, Feder, and Christensen (OFC)\cite{ofc} who proposed a cellular automaton model based on the BK model, 
that has quite remarkable properties
and is simple enough to be simulated very efficiently\cite{ofc2}. The OFC model considers a set of real valued variables 
$u_i$ where $i$ indicates the position in a two dimensional lattice. 
$u_i$ is the force that the substrate exerts on the block at position $i$, and it represents the
local stress between the sliding plates. 
The system is driven by uniformly increasing the values of $u_i$ with time at a rate $V$, simulating the tectonic loading of the plates. Every time one of the variables $u_i$ reaches 
a maximum pinning force (ordinarily set to an uniform, dimensionless value of 1), the local stress $u_i$ is relaxed by setting it to zero (thus the local stress drop $\Delta u$ is equal to 1). The local stress drop produces a stress increase onto neighbor sites according to $u_j \to u_j + \alpha$, where $j$ indicates a neighbor site to $i$. The value of $\alpha$
can vary between $0$ and $\alpha_c\equiv 1/z$, $z$ being the number of neighbors in the lattice. The case $\alpha=\alpha_c$ is called the conservative case, whereas $\alpha<\alpha_c$ are non-conservative cases. 
A discharge event can produce the overpassing of the maximum local stress on one or more than one neighbors, and in this way a large cascade can be generated. This cascade is called an event, and is identified with an individual earthquake (note that the full cascade is assumed to occur at constant time, namely, earthquakes are instantaneous). The size $S$ of an event is defined as the sum of all discharges $\Delta u$ that compose the event, and the magnitude is defined 
as $M=\frac{2}{3}\log_{10} S$, so to match (up to an additive constant) the usual definition used in geophysics \cite{scholz}.

The OFC model is typically simulated using open boundary conditions. In the case of periodic boundary conditions 
the model exhibits a strong global synchronization originated in the spatial homogeneity.
The OFC model displays an exponential decay of number of events as a function of magnitude compatible with a GR law. The $b$-value is however not universal, but depends on the value of $\alpha$. Realistic values of $b$ are obtained for $\alpha \simeq 0.2$ (with $z=4$). A cut-off at large event sizes exists that moves progressively to larger values for larger system sizes. 

After its introduction, the OFC model has been studied in great detail, trying to extract from it the characteristics that are observed in actual seismicity patterns. Although the finding of a GR decay law is a goal
of this kind of model, the spatial and temporal clustering of earthquakes observed in real seismicity are certainly not reproduced by the OFC model (see below the discussion about aftershocks in the OFC model), as well as they were not reproduced neither by the model of BK. I refer in particular to the phenomenon of aftershocks, that has a partial quantitative description through the empirical Omori law\cite{omori,utzu}. This law states that the number of earthquakes in excess of its average value after a large event decays as $\sim 1/(t+c)^p$, where $t$ is the time from the main shock, $p$ is an empirical exponent, and $c$ is a time constant in the range between minutes and hours.
Usually the value of $p$ is found to be close to one, although other values and even other functional forms have also been proposed \cite{bz2008}.

My contribution here is to modify the original OFC model in a way that allows for the existence of some kind of structural relaxation\cite{jagla}, or aging in the system. This modification produces the appearance of correlated events in the dynamical evolution, in particular aftershocks, generating earthquake sequences that qualitatively and quantitatively resemble real ones. In addition, the modified model will be shown to posses average friction properties that are compatible with those experimentally observed in rock friction studies\cite{marone}. In particular, I obtain the effect known as `velocity-weakening', namely, a reduction of the average friction force when the sliding velocity is increased, that is known to occur in rock friction, and plays a crucial effect in the triggering of earthquakes\cite{scholz}. Velocity weakening has been described phenomenologically in terms of the so called rate-and-state equations\cite{dieterich}, but no detailed quantitative theory exists for it.

In the next Section I introduce and justify the modifications that are made onto the OFC model. Results are presented in Section III. In Section IV, I show the dependence of some of the results on the kind of relaxation  mechanism used. Finally, In Section V, I give some qualitative interpretation of the appearance of aftershocks, summarize and conclude.

\section{The modifications to the model, and their justification}

Two modifications will be implemented onto the original OFC model. They are the existence of {\em random thresholds}, and {\em structural relaxation}. I now present and justify them separately.

1){\em Random thresholds:}
In the OFC model, the maximum values that the variables $u_i$ can withstand are set to a constant value of 1. Having in mind a realistic situation
of a heterogeneous fault, with the constitutive materials having different properties at different positions, it becomes natural to consider a case in  which the threshold values are not constant but have some spatial variation. 
In concrete, the values of the local thresholds will be called $u_i^{th}$,  and I draw them from a Gaussian distribution centered at $u_0$, with standard deviation $\sigma$. Each time $u_i$ overpasses the local threshold $u_i^{th}$, $u_i$ is updated to a new value. In concrete, I will use the update rule $u_i \to u_i-1$, i.e, I maintain (as in the original model) the prescription of a unitary local stress drop. Upon this drop of the local stress, the values of $u$ on neighbor sites are updated as before, namely $u_j \to u_j + \alpha$, for $j$ neighbor to $i$.  

Every time $u_i$ is updated, a new value is assigned to the local threshold $u_i^{th}$, taken from its original Gaussian distribution. This prescription is justified on the same physical arguments as before, since the sliding pieces can reasonably be thought to find different maximum strengths as sliding proceeds. I found that even a small value of $\sigma$
(about 0.05$u_0$) is sufficient to qualitatively modify the behavior of the OFC model (see results below). This means that the OFC model is not robust with respect to this perturbation, and since random variation of parameters is experimentally expected, the OFC model can probably be considered as an interesting dynamical system, but not as a realistic model of the seismic phenomenon. In the language of renormalization group theory, we can say that fluctuations in the threshold values are ``relevant" variables\cite{nota}.

2){\em Structural relaxation:} 
The existence of internal temporal effects in sliding systems is well established. Dieterich and Kilgore\cite{dk} were the first to provide direct evidence that solid bodies in contact experience plastic relaxation that induces the increase of real contact area over time. In turn, this increase of contact area is intimately related to the velocity weakening effect and thus it affects directly the characteristics of the seismic phenomenon\cite{scholz,persson}. Thus the second modification I make on the OFC model is the inclusion of
these relaxational processes\cite{notahainzl}. 
In order to justify the particular way in which relaxation will be introduced, I note that the plastic processes I am trying to model, always produce a reduction of the total energy $E$ stored in the system. This energy
will be dependent of the values of the $u_i$, namely there will be some functional form $E(u_i)$. 
The proposed relaxation mechanism causes a progressive reduction of this energy through a standard first order relaxation  equation of the form\cite{chaikin}

\begin{equation}
\frac{du_i}{dt} =R \left ( \nabla^2\frac{\delta E}{\delta u}\right )_i
\label{modelII}
\end{equation}   
where $\nabla ^2$ is the discrete Laplacian on the underlying square lattice, i.e., $(\nabla^2 f)_i= \sum_j f_j -4f_i$, where $j$ stands for the four neighbor sites to site $i$, and  lattice parameter is taken as the unit of length (see also the discussion in Section IV).

In order to have a concrete realization, we need to specify the form of the function $E(u_i)$. Whereas in principle this can be done by a detailed derivation of the OFC model from an elastic model, at present I will take the view of proposing the simplest form for the relaxation equation. This is obtained by taking $E\propto \sum_i u_i^2$, and it gives

\begin{equation}
\frac{du_i}{dt}=R \left(\nabla^2 u\right)_i +V
\label{us}
\end{equation}   
where the last term comes from the external driving, and a constant has been absorbed in the value of $R$. In the last section I will 
discuss the possibility of other analytical forms of the relaxation mechanism and the effect on the results obtained.

This mechanism\cite{jagla,jk} produces (for $V=0$) the progressive uniformization of the local forces $u_i$ on a time scale set by the relaxation parameter $R$.
Thus, the relevant parameter of the dynamics of the system will be the ratio $R/V$, that measures the competing effect between relaxation and the global driving.

Since I found that in the presence of non-uniform thresholds and for sufficiently large system sizes the results become independent of the boundary conditions used,
I chose to work always with periodic boundary conditions to reduce size effects as much as possible.
Note also that the dynamics of the model is independent of the average value of the thresholds $u_0$, since a change in this value produces only a rigid change of all $u_i$. I will formally take $u_0=1$, having in mind that a different value of $u_0$ can be considered if we want, for instance, to maintain
all $u_i$ variables to be positive at all times, as its physical interpretation would suggest. In addition, very small events (those that span ten or less lattice points) are systematically cut off from the results, since they are spuriously dependent on the underlying numerical lattice.

\section{Results}

\begin{figure}
\centerline{\includegraphics[width=.5\textwidth]{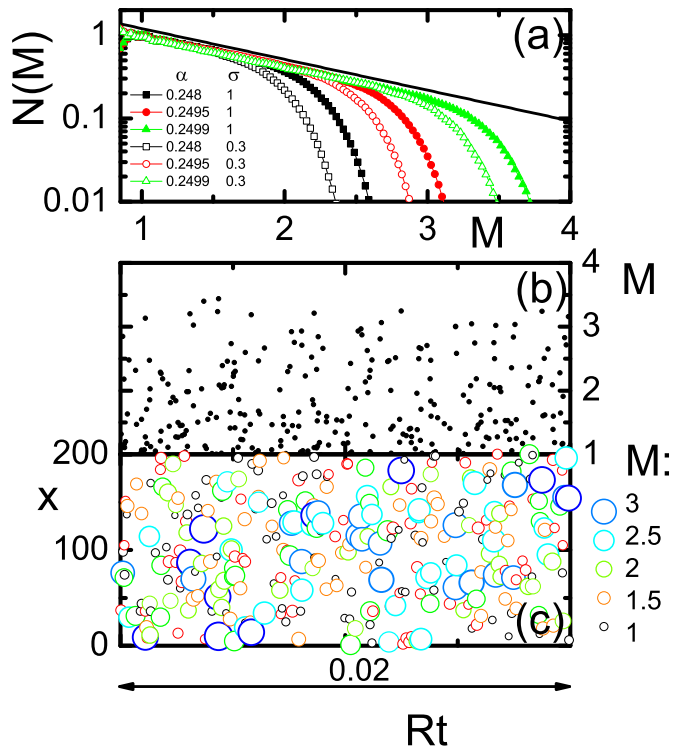}}
\caption{(Color online) (a) Magnitude-frequency distribution in the absence of relaxation ($R=0$), and for different parameters, as indicated. Continuous lines has a slope $b=0.37$, for reference. The large size cut-off is mainly controlled by $\alpha$, moving to infinity for $\alpha\to \alpha_c=0.25$. There are no important finite size effects in these results, as the system size ($L\times L=200\times$200) is much larger than the largest event that is observed to occur for each value of $\alpha$. (b)Magnitude vs time plot, 
and (c)projected position (along $x$-axis) vs time of epicenters of events with $M>0.9$, for the case  
for $\alpha=0.2499$, $\sigma=1$. Symbols size and color are magnitude dependent, according to the legend. No obvious sign of temporal or spatial correlation is observed.
}
\label{f1}
\end{figure}

Even in the case $R=0$, there are qualitative differences between the results obtained with the present model (that uses random thresholds) and with  the OFC model.
In Fig. \ref{f1} we see results for this case ($R=0$), for different values of $\alpha$ and $\sigma$. The decaying exponent of the
number of events with magnitude [Fig. \ref{f1}(a)] is
$b\simeq 0.37$, independently of the precise values of $\alpha$ and $\sigma$. There is a cut-off at large event sizes that moves towards infinity as $\alpha\to \alpha_c=0.25$. 
This is in contrast with the results in the OFC model, where the $b$ exponent depends on $\alpha$, and the cut-off depends on system size, and suggests that the physics of the model presented here is very different from that of the OFC model. Actually, the behavior I find for $R=0$ is consistent with the case of an elastic interface driven on top of a disordered pinning potential\cite{zapperi}. In particular the value of $b\simeq 0.37$ compares very well with the avalanche size exponent for an elastic
interface $\tau\simeq 1.25$  (as defined for instance in \cite{rosso2009}, note that $\tau=1+2b/3$).

An inspection of the 
spatial and temporal sequences of epicenters (i.e., the triggering positions) of the events presented in Fig. \ref{f1}(b)-(c), reveals no obvious correlations of any type.
The conclusion from here is that the model with random thresholds and without relaxation ($R=0$) is qualitatively different from the OFC model, but also far from being realistic in simulating
seismicity.

\begin{figure}
\centerline{\includegraphics[width=.5\textwidth]{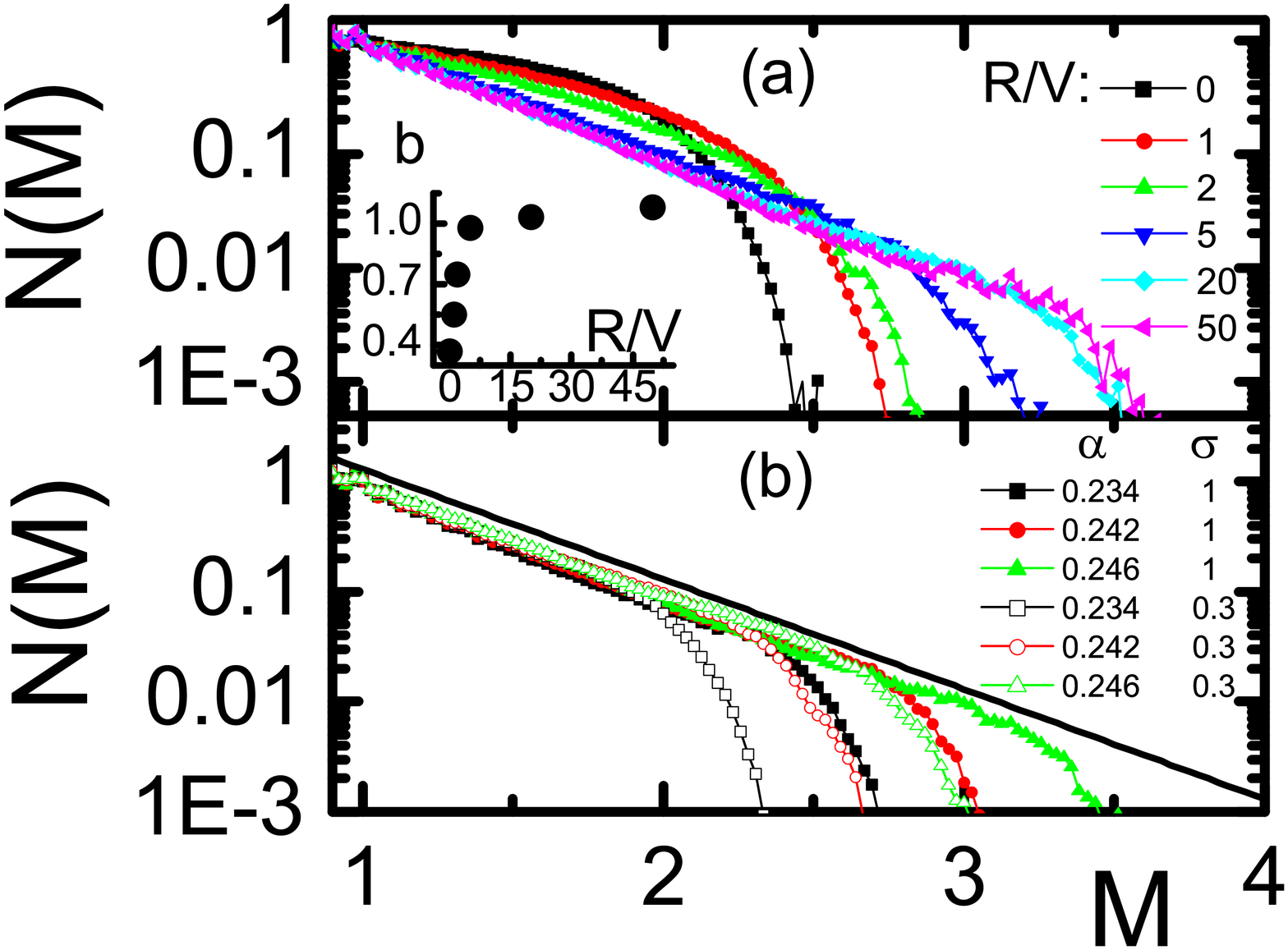}}
\caption{(Color online) (a) Magnitude-frequency distribution for increasing relaxation, and for 
$\alpha=0.246$, $\sigma=1$, and $L=200$. In the inset, the $b$ values extracted from the main panel are plotted as a function of $R/V$, showing a convergence to a well defined value for large $R/V$. Panel (b) shows the independence of the $b$ value on $\alpha$ and $\sigma$, and the progressive increase of the large size cutoff as $\alpha \to \alpha_c=0.25$.
}
\label{f2a}
\end{figure}

Qualitatively new results appear when $R$ is different from zero. The magnitude frequency distribution for increasing values of $R/V$ is shown in Fig. \ref{f2a}(a) for a fixed value of $\alpha=0.246$. As $R/V$ increases, the $b$ value increases [Fig. \ref{f2a}(a) inset]. Most remarkably, $b$ seems to reach a well defined value ($b \simeq 1.0$) when $R/V$ is large, that is independent of $\alpha$ and $\sigma$ (see Fig. \ref{f2a}(b)) and that is comparable to actual values observed in earthquakes. 
Fig. \ref{f2a}(b) also shows  that the large size cutoff of the GR plot is mainly governed by the value of $\alpha$, moving towards infinity as $\alpha \to 0.25$.
The appearance  
of a realistic $b$ value is encouraging since it is obtained without any tuning of parameters in the model (beyond the fact of $R/V$ being sufficiently large).

\begin{figure}
\centerline{\includegraphics[width=.5\textwidth]{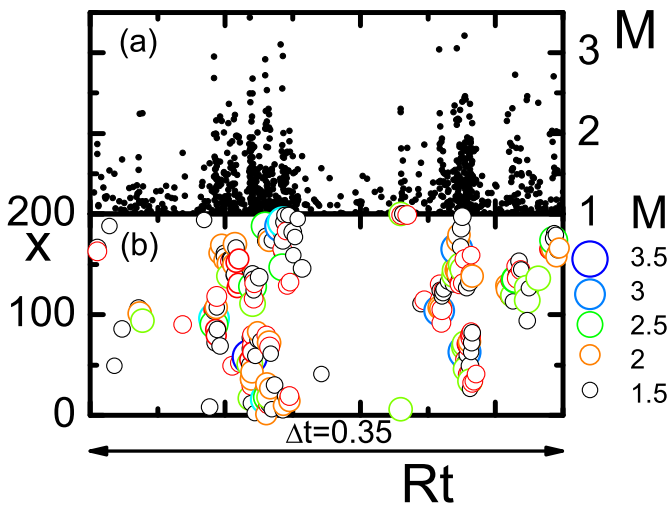}}
\caption{(Color online) 
(a) Magnitude-time plot and (b) projected position (along $x$-axis) vs time of epicenters of events with $M>1.5$, for the case $R/V=20$, $\alpha=0.246$, $\sigma=1$. Temporal and spatial clustering is apparent.
}
\label{f2b}
\end{figure}

\begin{figure}
\centerline{\includegraphics[width=.5\textwidth]{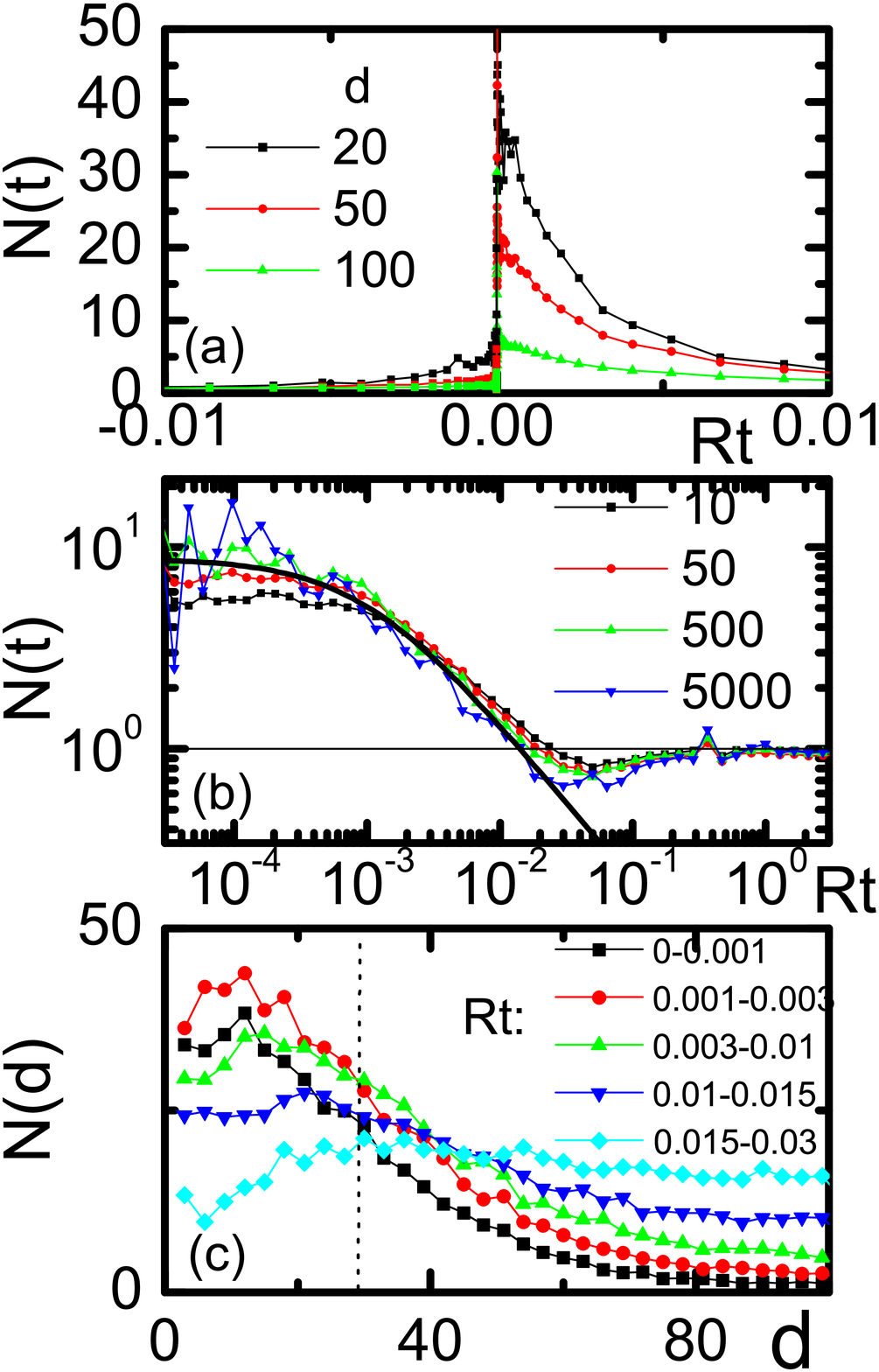}}
\caption{(Color online) (a) Aftershocks and foreshocks in the model with parameters $R/V=20$, $\alpha=0.244$, $\sigma=1$, system size 200 $\times$ 200. I show the accumulated histogram $N(t)$ of events occurring at time $t$ with respect to the large event (I consider about 200 large events, with magnitude $M>3.0$, with relative spatial distance of epicenters smaller than the cut off value $d$ . The histogram is normalized to the corresponding totally random situation [this means that $N(t\to \infty)=1$]. 
The accumulation of events following the large ones (aftershocks) is clearly visible. Foreshocks are observed but in a much
lesser extent.
(b) Activity in the whole system following main shocks, in logarithmic scale. 
For comparison, an Omori expression $a/(t+c)^p$ (with $a=0.035$, $c=0.001$, $p=0.8$) is also plotted. 
(c) Activity after main events, as a function of the distance to main shock epicenter $d$, for different time windows after the main event, as indicated. The typical radius of regions broken by the main shock is indicated by the vertical dotted line. A drift away from the main shock epicenter of later time aftershocks is clearly visible.
}
\label{omori}
\end{figure}

In addition to change the $b$-exponent of the magnitude-frequency distribution, relaxation generates non-trivial correlations in the spatial and temporal event distribution. 
We will see now that this clustering has features that are known to correspond to real earthquakes. 
First of all,
it is qualitatively seen in Fig. \ref{f2b} that events accumulate following large ones, and that the epicenters of the clustered events occur close to the epicenter of the large shock. This reproduces well known features of real aftershocks.
In order to provide a more quantitative characterization of aftershocks in the model, 
in Fig. \ref{omori}(a), I show the result of calculate histograms of events occurrence
around main shocks.
For this analysis, a main shock is operationally defined as any event having $M>3.0$ (this corresponds to events producing a rupture region of linear size about 60 lattice sites). Time is set to zero at the main shock. Different curves are presented, that correspond to events occurring within a given spatial distance $d$ from the main shock epicenter. Curves are normalized in such a way that $N(t\to \infty)=1$.
The over abundance of events following large ones is clear. There is also some over abundance of events preceding large ones (foreshocks), but in a much lesser extent than the case of aftershocks. 
In order to compare with an Omori expression, in Fig. \ref{omori}(b) I plot the evolution of the activity after the main shock, in logarithmic scale, and using different lower cutoff values to define aftershocks.
Curves are compatible with an Omori law, but we see that the activity soon becomes masked by the background activity,
rendering a quantitative determination of parameters in the Omori expression very unconstrained.
(below we will see a trick to avoid this problem). In addition, we see that the convergence towards the background activity is not monotonous, instead a time window of lower-than-average activity (at times around $0.05/R$) is observed.

In order to analyze quantitatively the spatial clustering of aftershocks, in Fig. \ref{omori}(c) I present histograms of activity in the system for different time windows after the main shock as a function of the distance to the main shock epicenter. It is observed that in very short times after the main shock, aftershocks occur rather uniformly in a region of size comparable with the rupture region of the main shock, and fewer events are observed at larger distances. When we consider later aftershocks, the spatial distribution clearly drifts away of the main shock epicenter. This is in nice agreement with the observed behavior of actual seismicity patterns. The lower-than-average
activity region indicated in (b) is seen here [latest curve in (c)] to be due to the eventual appearance of a region of depleted activity (with respect to its spatial average) close to the main shock epicenter.

\begin{figure}
\centerline{\includegraphics[width=.5\textwidth]{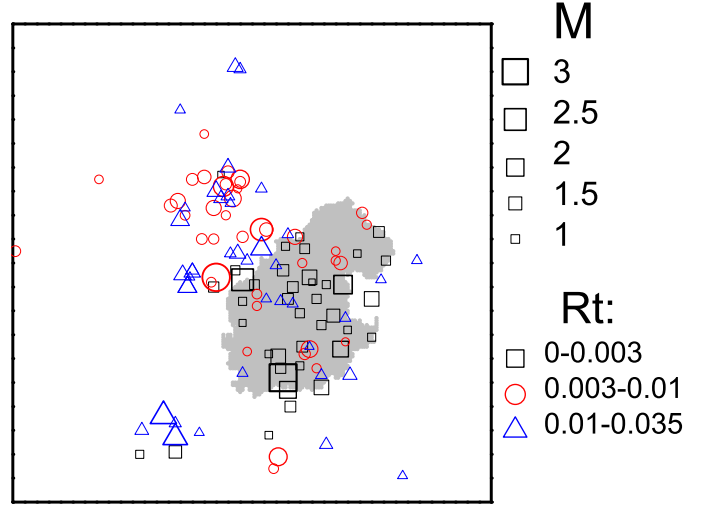}}
\caption{(Color online) In gray, the broken region of a particular main shock. Symbols are the epicenters of events following the main shock. Symbol size depends on magnitude, and symbol type indicates different time windows after the main shock. The figure displays a system of size 200 $\times$ 200, and parameters are $\alpha=0.244$, $R/V=20$, and $\sigma=1$.
}
\label{omori2}
\end{figure}

A complementary, more visual example of aftershocks spatial distribution is presented in Fig. \ref{omori2}. There I show the actual region that was broken by a particular main shock, and the activity following this event. Squares, circles, and triangles correspond to three time windows of progressively later events, as indicated in the legend ($t=0$ at the main shock). Size of the symbol increases with magnitude. In this example we see again that aftershock activity starts mainly within the region broken by the main shock, and then progressively drifts away.

\begin{figure}
\centerline{\includegraphics[width=.5\textwidth]{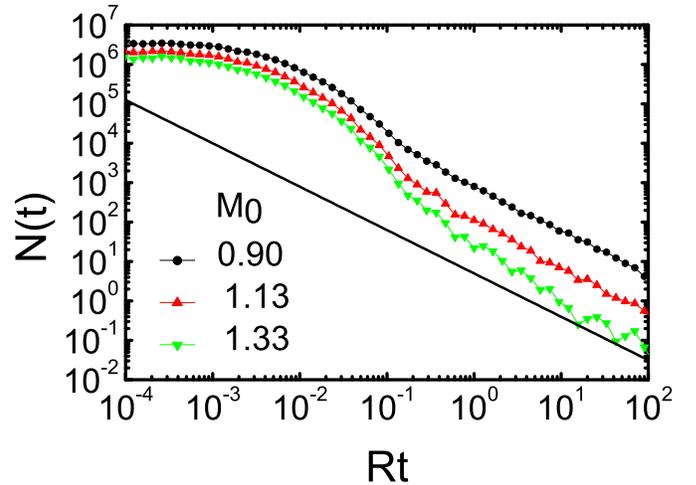}}
\caption{(Color online) Aftershock decay rate in simulations in which tectonic loading is stopped ($V=0$) after a main shock has occurred (parameters as in Fig. \ref{omori}). The asymptotic form of the decay follows nicely an Omori decay (for reference, the continuous lines has a slope of 1.1). Different curves correspond to different lower cutoff values $M_0$ used to count aftershocks.
}
\label{f4}
\end{figure}

In the previous analysis, the aftershock statistics is limited by the fact that tectonic loading continues to trigger events in the system that rapidly mask the aftershock tail of previous main shocks. 
However a numerical trick can be implement to overcome this problem in order to study aftershocks in more detail. Along a simulation, once a main shock is detected I stop tectonic loading (i.e. setting $V=0$) and look for the aftershock occurrence. The process is repeated many times to accumulate good statistics. The result is shown in Fig. \ref{f4}. Now we can follow the aftershock decay rate for few orders of magnitude in time. It is seen that a very good power law (Omori like) decay is obtained with a $p$ exponent around 1.1.  I notice however that at short times after the main shock some over abundance of events with respect to the asymptotic rate is observed. This can be interpreted as a larger $p$-value if the aftershock sequence is observed in a limited time interval.

It is necessary to mention here that aftershocks of a very peculiar type have been found in the original OFC model\cite{ofc-as}. In my view, these aftershocks reflect once more the kind of synchronization that the OFC model is prone to, and have nothing to do with aftershocks observed in real seismicity. In particular, they completely disappear once a randomness in the thresholds of about $5\%$ is included\cite{jagla_unp}, indicating that they are not a robust property. Also, their existence depends
exclusively on external loading, i.e., if at some moment the external driving vanishes, all seismic activity ceases immediately. Namely, the plot equivalent to that in Fig. \ref{f4} for the OFC model would be completely void.

\begin{figure}
\centerline{\includegraphics[width=.5\textwidth]{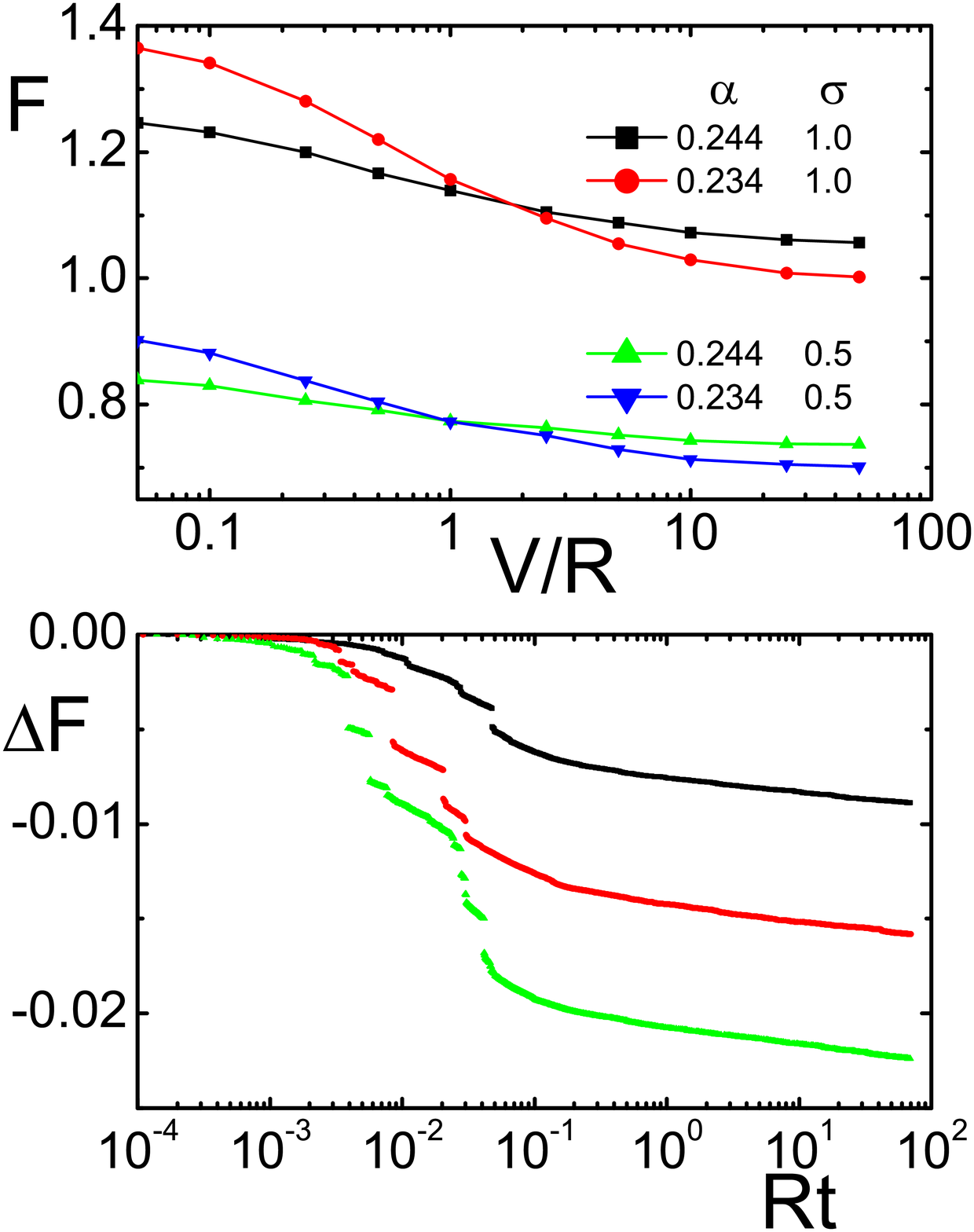}}
\caption{(Color online) (a) Average friction force $F$ as a function of velocity in a system of size $200\times 200$ sites and other parameters as indicated. Velocity weakening following an approximately logarithmic dependence on velocity is clearly observed. (b) Results of three individual realizations for the stress decay in a system of size $200\times 200$ ($R/V=20$, $\alpha=0.244$, $\sigma=1$)
after the abrupt stopping of loading at some arbitrary time (set to zero). 
}
\label{f5}
\end{figure}

In addition to generate a realistic $b$ value, and produce clustering effects, structural relaxation generates global frictional properties that are comparable to what is observed in laboratory experiments of friction between solids. I refer in particular to the so-called velocity weakening properties of the friction phenomenon\cite{scholz,marone}, and in general to the  phenomenology given by rate-and-state equations\cite{dieterich}, widely used in seismological analysis. Velocity weakening means that the average friction force $F$ between the sliding bodies decreases as a function of relative sliding velocity. In the BK model, this behavior has to be introduced by hand in the form of a tailored friction law between blocks and substrate.
The OFC model, on the other hand, can be considered to generate a friction force (that is obtained in this case as the spatial and temporal average of the local friction forces $u_i$) that is independent of sliding velocity, since sliding velocity plays no role in the dynamics of the system, as earthquakes are assumed to be instantaneous. But in the present model, the interplay between structural relaxation and the external driving velocity generates 
a friction force $F$ that depends on velocity. This dependence is of the 
velocity-weakening type, as can be seen in the results in Fig. \ref{f5}(a).
The decrease of friction force with velocity is mainly logarithmic in about three orders of magnitude of velocity variation, quite comparable to the experimental results in Ref. \cite{marone}. The increase of friction force as velocity is reduced can be qualitatively understood if we consider that at lower velocities, the system has more time to reach more stable configurations (with lower energies). Thus, larger forces have to be applied  in order to, eventually, take out the system from these configurations, and this means a larger friction force.

Another effect that the present model reproduces is the stress relaxation that is observed after 
loading is stopped in laboratory friction experiments \cite{marone}. We have already seen that in the present model activity decays slowly after tectonic loading is stopped, and this produces a stress relaxation that follows an almost logarithmic trend. In fact, in Fig. \ref{f5}(b) we see examples of the stress decay after stopping loading, where the effect of individual large aftershocks is seen as abrupt stress drops.

\section{Dependences on the relaxation mechanism}

The use of a conserving form of Eq. (\ref{modelII}), i.e., the inclusion of the Laplacian operator, instead of
a non-conserving form of the kind
${du}/{dt} =-\lambda \delta E/{\delta u}$ is difficult to justify on first principles.
One possible {\em a posteriori} justification is that a non-conserving dynamics produces, in the absence of tectonic loading,
the evolution of the system towards a state with $u_i=0$ everywhere, i.e., a stress free state. In particular, the almost logarithmic stress decay obtained in the last Section would not occur, instead we would observe an exponential decay towards zero stress. This is not a realistic situation for the present problem, although it could possibly be appropriate to model a viscoelastic response. The necessary requirement to have a system that is able to maintain a constant stress under static conditions is that relaxation does not modify the spatial average $\bar u$ of $u_i$. Eq. (\ref{us}) certainly satisfies this requirement, but other forms are possible. I compare in this Section some of the results obtained using Eq. (\ref{us}) with two other possibilities for the relaxation mechanism, namely

\begin{eqnarray}
\frac{du_i}{dt}=-R (u_i-\bar u) +V\label{us2}\\
\frac{du_i}{dt}=-R \left(\nabla^4 u\right)_i +V\label{us3}.
\end{eqnarray}
(I use the same symbol $R$ for the relaxation parameter in all cases, although its numerical value may differ).
Both Eqs. (\ref{us2}) and (\ref{us3}) do not modify the spatial average $\bar u$ of the $u_i$ variables. Eq. (\ref{us2}) is a sort of ``mean field" implementation of the relaxation mechanism. It is not a realistic possibility in a physical system with local interactions, but is an interesting case of study to compare with. Eq. (\ref{us3}) incorporates a double Laplacian to drive the temporal variations of $u_i$. 
I will consider this to be simply another formal possibility, although it can be given a more physical justification by deriving the present model as a limit of an elastic spring block-model in the presence of relaxation (see Ref. \cite{jk}). Now I will present a comparison of the results obtained using Eqs. (\ref{us}), (\ref{us2}), and (\ref{us3}).

\begin{figure}
\centerline{\includegraphics[width=.5\textwidth]{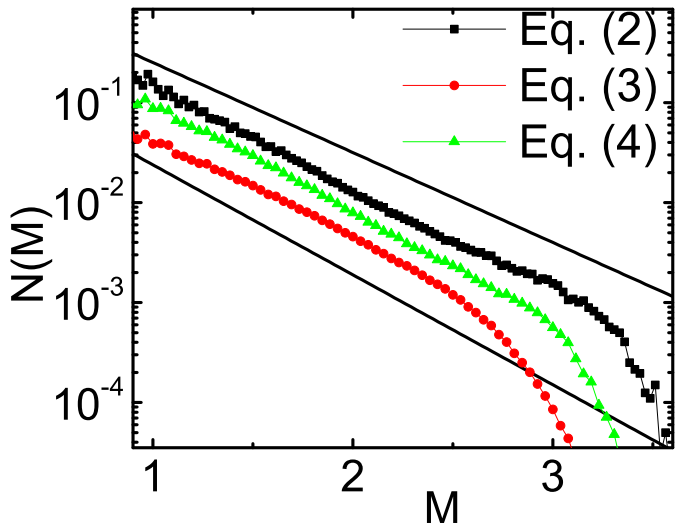}}
\caption{(Color online) Magnitude frequency distribution obtained with the three different relaxation mechanisms proposed in text, for $R/V=20$, $\alpha=0.246$, and $\sigma=1$ in all cases (curves are not normalized, and were vertically displaced for better comparison). The value of $b$ for the three mechanisms is within a range of $0.9<b<1.1$, close to experimentally observed values. The two limiting slopes are plotted as continuous lines, for reference.
}
\label{extra1}
\end{figure}

First of all, the three implementations produce a modification of the $b$ exponent of the GR decay, that tends to a conserved value when $R/V$ is large enough, in the three cases. The conserved value, is within the range 0.9-1.1 in the three cases. This can be seen in Fig. \ref{extra1}, where I present the magnitude frequency distribution for the three mechanisms, for a rather large value of $R/V$ (so the $b$ value has already reached its asymptotic value). 

\begin{figure}
\centerline{\includegraphics[width=.5\textwidth]{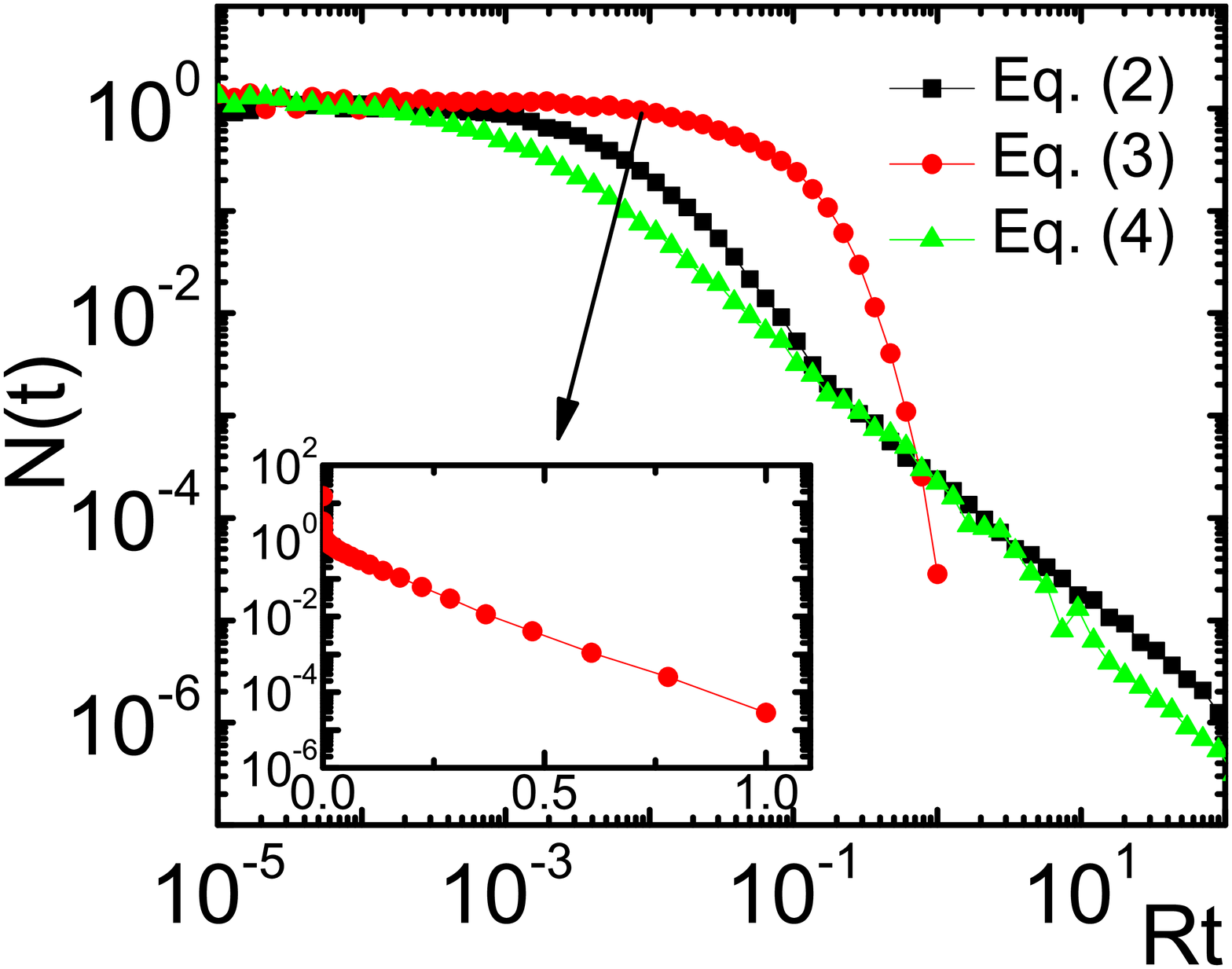}}
\caption{(Color online) Time decay of aftershocks in a 200 $\times$ 200 system in the case in which loading is stopped after a main shock is detected. 
The three curves correspond to the three different realizations of the relaxation mechanism. The curve corresponding to Eq. (3) is plotted in a different scale in the inset, to highlight an exponential decay of aftershocks in this case.
}
\label{extra2}
\end{figure}

The other comparison refers to the asymptotic decay of aftershocks. In Fig. \ref{extra2} we observe the aftershock activity (using the prescription of stopping tectonic loading) after main shocks. There is a clear difference in aftershock behavior for the mean field case of Eq. (\ref{us2}). In this case, the aftershocks activity decays exponentially with time. 
On the other hand, the difference between the behaviors using Eq. (\ref{us}) and (\ref{us3}) is more subtle. I already indicated that using Eq. \ref{us} an over-abundance of aftershocks at small times is observed, before an Omori decay (with $p\simeq 1.1$) is observed. For the bi-Laplacian relaxation the aftershock excess is much less pronounced, and an almost perfect Omori law  with $p\simeq 1.25$ is observed in this case. 

The results in this Section indicate that some qualitative features of the model are robust upon a change of the relaxation mechanism, although differences in the details can be expected.

\section{Further analysis and conclusions}

\begin{figure}
\centerline{\includegraphics[width=.5\textwidth]{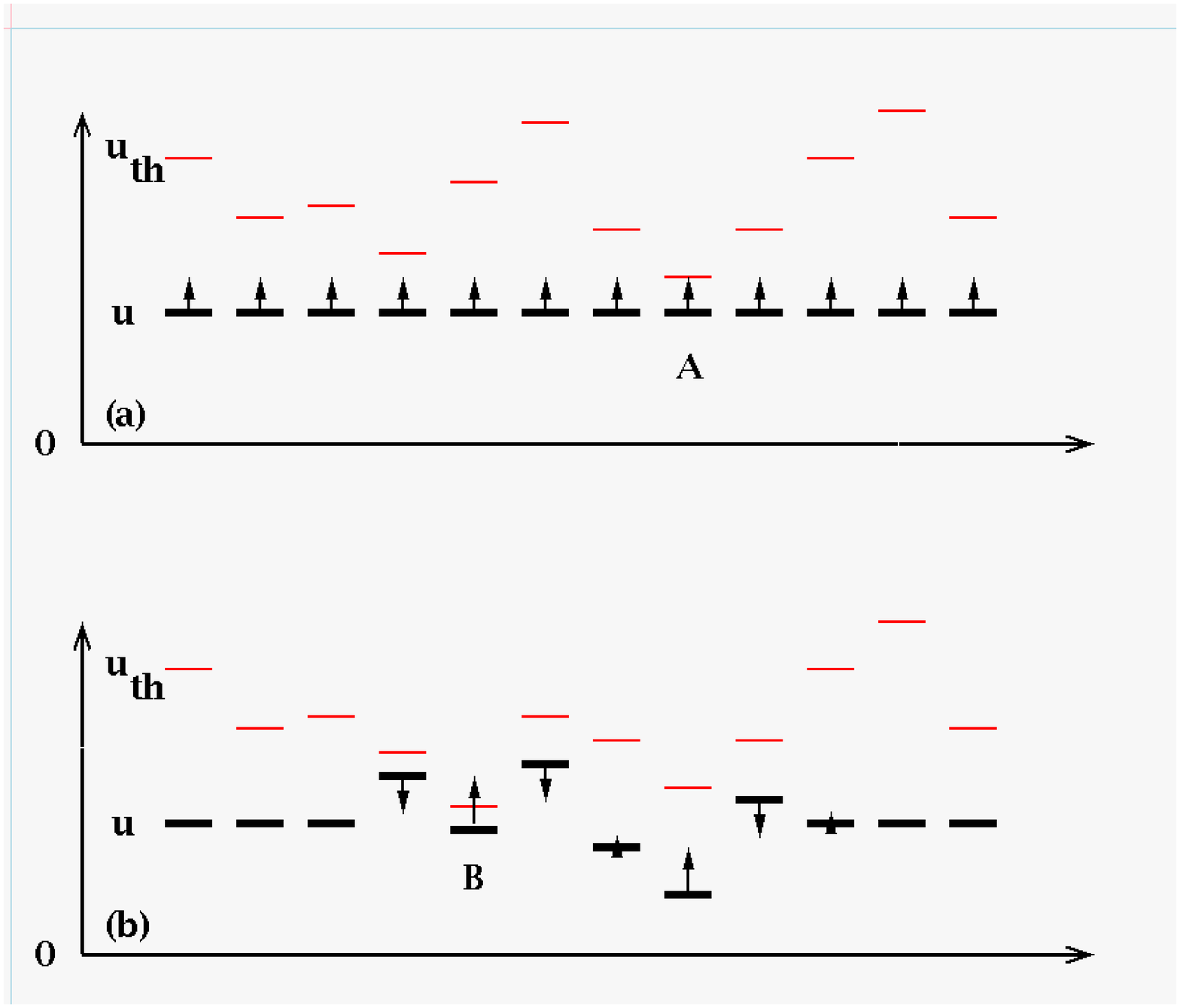}}
\caption{(Color online) Pictorial description of main shocks and aftershocks in a one dimensional version of the model, in the limit $V/R\to 0$. Vertical position of thick (thin) segments are the values of $u_i$ ($u_i^{th}$). Rate of change of $u$'s is indicated by the arrows. In (a), a relaxed (thus uniform) distribution of $u$'s increases at a rate $V$. At the point indicated by $A$, the first threshold will be overpassed and a main shock will be triggered. After the main event, the system reaches a situation depicted in (b). Relaxation tends to make the values of $u$ uniform across the sample, and in this process an aftershock (in this case at point $B$) can be triggered.
}
\label{f6}
\end{figure}

Finally, in order to clarify the origin of aftershocks in the model, we can make an analysis of the limiting case in which
$V/R\to 0$. This case can be realized in the following way. We set temporarily $V$ to zero, and allow only the evolution given by the relaxation term in Eq. (\ref{us}). This evolution is continued until we can guarantee that no other event will be triggered by the relaxation alone. At this point the values of $u_i$ can be set everywhere equal to their mean value, and this flat interface can be driven by the external velocity $V$ until a new instability occurs. In this limit, a precise distinction between main shocks and aftershocks can be given: aftershocks are events that are triggered by the term proportional to $R$ in Eq. (\ref{us}), while main shocks are triggered by $V$ (note that in this case, main shocks are not defined in terms of their intensity, in fact, it can occur that a main shock produces an aftershock
of larger magnitude that the starting event).  The mechanism of aftershock triggering is illustrated in this limit in Fig. \ref{f6}. In order for the relaxation to be able to trigger events by itself,
the thresholds cannot be uniform, since in that case, starting with $u_i<1$
after a given event, 
evolution through Eq. (\ref{us}) with $V=0$ cannot produce any $u_i>1$. However, if thresholds have some randomness,
the evolution according to Eq. (\ref{us}) with $V=0$ can produce
$u_i>u_i^{th}$ at some position (particularly at those
with the smallest thresholds), and an aftershock is triggered. This highlights the crucial role played by a non-uniform distribution of thresholds in the appearance of aftershocks. It is thus not surprising that aftershocks are observed only if the distribution of thresholds has a dispersion $\sigma$ larger than some minimum value that turns out to be about 0.25.

Summarizing, I have presented a model that is based on the one proposed by Olami, Feder, and Christensen\cite{ofc} to study the dynamical appearance of slip events between tectonic plates. Modifications consider the existence of random thresholds, and the possibility of relaxation in the system. Relaxation acts trying to strengthen the contact between the sliding surfaces if they remain at rest relative to each other. When the surfaces slip, the contacts refreshen and there is a competence between the relaxation mechanism and the external driving of the system. With this kind of modification I have been able to generate earthquake sequences that contain many of the features observed in real seismicity. In particular, I have presented results of temporal clustering of events following a main shock according to the Omori law, and spatial clustering of these events around the epicenter of the main shock. Also, an appropriate decay of number of events as a function of magnitude according to the Gutenberg-Richter law, with a realistic $b$-exponent has been obtained, and the value of $b$ was shown to be independent of other parameters of the model, relaxation assumed to be sufficiently large.
Although the present model does not introduce velocity weakening directly, this effect appears as a consequence of structural relaxation. In this way, the model gives also as a physical basis for the rate-and-state equations used to describe frictional properties of solids.

\section{Acknowledgments}

This research was financially supported by Consejo Nacional de Investigaciones Cient\'{\i}ficas y T\'ecnicas (CONICET), Argentina. Partial support from
grant PICT 32859/2005 (ANPCyT, Argentina) is also acknowledged.

\newpage


\begin{thebibliography}{4}
\bibitem{nota}Note that if the inclusion of this randomness did not produce any qualitative change in the behavior of the system, then it would have been sensible to work in the case $\sigma=0$.
\end{thebibliography}
\end{document}